\documentclass[oneside,english]{amsart}
\usepackage[T1]{fontenc}
\usepackage[latin1]{inputenc}
\usepackage{amsthm}

\makeatletter
\numberwithin{equation}{section}
\numberwithin{figure}{section}
\theoremstyle{plain}
\newtheorem{thm}{Theorem}
  \theoremstyle{plain}
  \newtheorem{prop}[thm]{Proposition}

\usepackage{graphicx}
\usepackage{babel}
\theoremstyle{plain}

\theoremstyle{plain}
\usepackage{babel}

\makeatother

\begin{document}
\makeatother

\title{One-dimensional mechanical networks and crystals}

\author{V. A. Malyshev }

\maketitle

\section{Introduction}

Thermal expansion and Hooke's law are among the fundamental properties
of macroscopic matter. In many cases it has satisfactory microscopic
explanation in the framework of equilibrium statistical mechanics.
This is a standard fact for the thermal expansion of gases and liquids.
Derivation of linear theory of elasticity modules in general translation-invariant
case see in \cite{BavChoqFon,Bav}.

However, for crystals, where there is no translation invariance, the
corresponding rigorous theory does not still exist. The nonstandard
character of this problem is related, as we explain below, to the
coordination between local and global coordinate systems. The goal
of this paper is to consider the simplest natural one-dimensional
model in the global coordinate system. This model can be considered
as one-dimensional crystal or one-dimensional network (macromolecules,
microtubules etc.) in the biological cell, see \cite{Boal}.

Standard courses of physics, for example \cite{FeyLeiSan2,Giri,LanLif7},
or even more specialized books, for example \cite{Weiner,Boal} try
to expalin thermal and elastic expansions of srystals via the oscillation
of atoms about their equilibrium positions in the crystal lattice.
In particular, the so called harmonic approximation and its nonlinear
analogs are normally used. Any atom is considered to be in the vicinity
of the equilibrium point. This is quite natural in the (microscopically)
local coordinate system. However if we assume this for ANY atom in
global coordinate system, then we cannot get any macroscopic expansion. 

The main idea is that the system is swelling uniformly at each point,
and in any global coordinate system almost atoms become macroscopically
far from their equilibrium positions. Note that similar situation
was described in the so called random grammars, see \cite{Mal01}.
Technically, this is achieved with another idea: to give atoms a possibility
to be far away from their equilibrium position one should impose some
restrictions on their motion. The simplest restriction would be to
disallow particles jump through each other, that is the order of particles
should be fixed. This idea of restricting the configuration space
is close to one discussed by O. Penrose \cite{Pen} in more complicated
two and three dimensional cases. He argued that restrictions on the
configuration space are necessary to get dependence of the free energy
not only on the volume but also on its form. In our one-dimensional
model this idea finds its rigorous justification together with very
precise calculations.

It is important to note that due to our restrictions one can give
sense to the partition function of finite number of particles in the
infinite volume. 

We consider the system of $N+1$ particles (molecules, atoms) on the
real line \[
0=x_{0}(t)<...<x_{k}(t)<...<x_{N}(t)\]
 where $t\geq0$. The dynamics of this system is defined via smooth
symmetric two-particle potential $V(x_{i}-x_{j})$\.{ } We make the
following assumptions concerning this potential:
\begin{enumerate}
\item For $u>0$ the potential $V(u)$ has a unique minimum at some point
$a>0$. Moreover, this minimum is quadratic;
\item We assume that\[
V(0)=\infty\]
 and make some comments on this assumption. In more realistic situation,
when the real line is, for example, the $x$-axis in $R^{d}$, the
particles can deviate from this axis in perpendicular directions,
and can in principle pass through each other in the $x$-direction.
However time scale for such transitions is slower than their movement
without changing the order. Thus our assumption is a stable approximation
to the realistic metastable situation. 
\item The strongest technical assumption is that is we consider only nearest
neighbor interaction, that is the total energy is\[
\sum_{k=1}^{N}V(x_{k}-x_{k-1})\]
 This is quite natural for the one-dimensional networks but less natural
for crystal models. From this assumption it follows, in particular,
that there exists a unique state with minimal energy (ground state)\[
x_{k}=ka\]
 Without this assumption the situation with periodic ground states
is more complicated, see \cite{Radin1,Radin2}, but we will not need
this; 
\item Our last assumption is that\[
V(u)\rightarrow\infty\]
 sufficiently fast (it is sufficient like $\frac{1}{u^{n}},n>1$)
as $u\rightarrow\infty$\.{ } This assumption is necessary to make
the partition function, for finite number of particles on the real
line, finite. This is also a stable approximation to the metastable
situation - to make impossible breaking the chain of particles into
separate parts. 
\end{enumerate}

\section{Thermal expansion}

Assume that the left coordinate is fixed $x_{0}=0$ and put

\[
u_{i}=x_{i}-x_{i-1}>0,i=1,...,N\]
 One can write the Gibbs density of the vector $(u_{1},...,u_{N})$
for temperature $T=\beta^{-1}$ as the following expression, which
allows factorization, \[
Z_{N}^{-1}exp(-\beta\sum_{i=1}^{N}V(u_{i}))dx_{1}...dx_{N}=Z_{N}^{-1}\prod_{i=1}^{N}exp(-\beta V(u_{i}))du_{i}\]

It follows that the random variables $u_{i}$ are independent, identically
distributed, and do not depend on $N$. This makes the thermodynamic
limit as $N\rightarrow\infty$ trivial. In particular the means $<u_{i}>$
do not depend on $N,i$ and can be written as\begin{equation}
<u_{i}>=m(T)=\frac{\int_{0}^{\infty}uexp(-\beta V(u)du}{\int_{0}^{\infty}exp(-\beta V(u)du}\label{fraction}\end{equation}
 Then the total length of the chain is\[
<x_{N}>=Nm(T)\]
 We will consider here two cases: low temperatures, that is small
deviations from the ground state and small perturbations around the
state with fixed temperature. We will find conditions on the potential
when macroscopic thermal expansion holds.

\subsubsection{Low temperatures}

Assume that the potential has the following Taylor expansion at the
point $x=a$\[
V(a+y)=V(a)+c_{2}y^{2}+c_{3}y^{3}+c_{4}y^{4}+o(y^{4}),c_{2}>0\]

\begin{thm}
For low temperatures $T=\beta^{-1}$

\[
m(T)=a+m_{1}T+o(T)\]
 where we will call $m_{1}$ the coefficient of thermal expansion,
If $c_{3}\neq0$, then\[
m_{1}=-\frac{3c_{3}}{4c_{2}^{2}}\]
 This coefficient is positive iff $c_{3}<0$. If on the contrary $c_{3}=0,c_{4}\neq0$,
then

\[
m_{1}=0\]
\end{thm}
\begin{description}
\item [{Remark}] If in the vicinity of $a$ the potential is purely quadratic,
that is it equals $(x-a)^{2}$, then also $m_{1}=0$. 
\end{description}
The physical example of the case $c_{3}<0$ can be the Lennard-Jones
potential (except of the sufficiently large values of the argument)\[
V_{LJ}=(\frac{\sigma}{r})^{12}-(\frac{\sigma}{r})^{6}\]
 where (for $\sigma=1$) at the minimum $a=\sqrt[6]{2}$ we have\[
c_{3}=-\frac{63}{\sqrt{2}}\]

Proof. Consider the asymptotic expansion (for large $\beta$) of the
numerator and the denominator in (\ref{fraction}). The standard Laplace
method can be used. Then\[
\int_{0}^{\infty}exp(-\beta V(u)du=\sqrt{\frac{\pi}{\beta}}exp(-\beta V(a))(b_{0}+b_{2}(2\beta)^{-1}+O(\beta^{-2}))\]
 \[
\int_{0}^{\infty}uexp(-\beta V(u)du=\sqrt{\frac{\pi}{\beta}}exp(-\beta V(a))(d_{0}+d_{2}(2\beta)^{-1}+O(\beta^{-2}))\]
 The coefficients can be found from the equations\begin{equation}
\chi'(z)=\sum_{k=0}^{\infty}b_{k}z^{k}\label{coef1}\end{equation}
 or correspondingly\begin{equation}
(a+\chi(z))\chi'(z)=\sum_{k=0}^{\infty}d_{k}z^{k}\label{coef2}\end{equation}
 where the function $\chi$, in the vicinity of the point $z=0$,
is defined by the conditions\begin{equation}
V(a+\chi(z))=V(a)+z^{2},\chi(0)=0,\chi'(0)>0\label{zamena}\end{equation}
 see theorem 1.2.4 in \cite{Evgr}. Put\[
\chi(z)=f_{1}z+f_{2}z^{2}+...\]
 Then by (\ref{zamena}) \[
f_{1}=c_{2}^{-\frac{1}{2}},f_{2}=-\frac{c_{3}}{2c_{2}^{2}},\]
 \[
f_{3}=-\frac{c_{2}f_{2}^{2}+3c_{3}f_{1}^{2}f_{2}+c_{4}f_{1}^{4}}{2f_{1}}=\]
 Then from (\ref{coef1}),(\ref{coef2}) we get\[
b_{0}=f_{1},b_{2}=3f_{3}\]
 \[
d_{0}=af_{1},d_{2}=3(f_{1}f_{2}+af_{3})\]

That is why\[
d_{0}b_{0}^{-1}=a\]
 and\[
m_{1}=\frac{d_{2}}{2b_{0}}-\frac{d_{0}b_{2}}{2b_{0}^{2}}=\frac{d_{2}}{2b_{0}}-\frac{ab_{2}}{2b_{0}}=\frac{3}{2}f_{2}\]
 The theorem follows.

\subsubsection{Small perturbations}

One can write\[
m(\frac{1}{\beta-\epsilon})=m(T=\beta^{-1})+m_{1}\epsilon+o(\epsilon)\]
 and the differentiation gives\[
m_{1}=<uV>_{\beta}-<u>_{\beta}<V>_{\beta}\]
 It follows that the criterion of the positive expansion is the positivity
of the correlation between random variables $u$ and $V$. Again,
if $c_{3}\neq0$, then for large $\beta$ the latter correltion is
positive iff $c_{3}<0$.

\section{Hooke's law}

Hooke's law for the change $\Delta l$ of the length of cylindrical
bar under the action of the force $F$ is the following\[
\frac{F}{S}=\kappa\frac{\Delta l}{l}\]
 where $S$ - the area of the cylinder crossection, $l$ - is the
length of the bar.

Under the previous conditions assume also that there is an external
potential $V_{ext}(x)=-Fx$, acting on the rightmost particle $x_{N}$.
Remind that $x_{0}=0$ is assumed to be fixed. Then\[
exp(-\beta\sum_{i=1}^{N}V(u_{i})-\beta V_{ext}(x_{N}))=exp(-\beta\sum_{i=1}^{N}(V(u_{i})-Fu_{i}))\]
 as $x_{N}=u_{1}+...+u_{N}$. Define\[
m(T,F)=\frac{\int_{0}^{\infty}uexp(-\beta(V(u)-Fu))du}{\int_{0}^{\infty}exp(-\beta(V(u)-Fu))du}\]
Then the elastic expansion is\[
m(T,F)-m(T,0)\]

\begin{thm}
For fixed temperature $T$, sufficiently small $F$ and any potential
$V$, satisfying the conditions in the introduction, the elastic expansion
is positive for positive $F$ and negative for negative $F$. Moreover,

\[
m(T,F)-m(T,0)=RF+o(F)\]
 where the elastic modulus\begin{equation}
R=\frac{\partial m(T,F)}{\partial F}|_{F=0}=\beta(<u^{2}>-<u>^{2})\label{modulus}\end{equation}
 .
\end{thm}
Proof. The formula (\ref{modulus}) is proved by simple differentiation.
It is sufficient to observe now that the right hand side of (\ref{modulus})
is the variance of nonzero random variable.

\section{Remark about the absence of expansion for oscillations around ground
state equilibrium}

In the standard courses of physics normally the so called harmonic
approximation is used with the hamiltonian (see for example \cite{Giri},
section 2.14)\[
\sum_{k,m=1}^{N}C_{km}\xi_{k}\xi_{m}+\sum_{k=1}^{N}\frac{1}{2}(\frac{d\xi_{k}}{dt})^{2}\]
 with symmetric matrix $C_{km}$, where $\xi_{i}=x_{i}-ia$ are the
deviations from ground state coordinates. There is however an evident
statement concerning joint distribution of the random variables $\xi_{i}$.
\begin{prop}
Let the distributions of $\xi_{i}(t)$ are such that for any $t,i$
and any sufficiently large $x$

\[
P(|\xi_{i}(t)|>x)=o(x^{-1})\]
 Then for $N\rightarrow\infty$ and any $t$ we have

\[
X(t)=max_{i}x_{i}(t)-min_{i}x_{i}(t)\sim Na\]

\end{prop}
In fact, for any $\epsilon>0$ the union of the events\[
A_{i}=\{|\xi_{i}(t)|>\epsilon N\}\]
 has the probability which tends to zero as $N\rightarrow\infty$.

Thus, one cannot get any macroscopic expansion in this way.

\end{document}